\documentclass[useAMS,usenatbib]{mn2e}

\usepackage{times}
\usepackage{graphicx}
\usepackage{rotating}

\def\pcm3{{\rm\thinspace cm^{-3}}}

\newcommand{\kms}{\mbox{$\rm km\thinspace\s^{-1}$}}

\def\n_h{{\rm n_{H}}}

\def\NH1{{$N_{\rm HI}~$}}

\def\s{\ \ \ }          

\def\ga{{\rm\thinspace gauss}}

\def\s{{\rm\thinspace s}}

\def\approxlt{\mathrel{\hbox{\rlap{\lower .5ex \hbox {$\sim$}}
        \raise .15 ex \hbox{$<$}}}}
\def\approxgt{\mathrel{\hbox{\rlap{\lower .5ex \hbox {$\sim$}}
        \raise .15 ex \hbox{$>$}}}}

\def\la{\mathrel{\hbox{\rlap{\hbox{\lower4pt\hbox{$\sim$}}}\hbox{$<$}}}}
\def\ga{\mathrel{\hbox{\rlap{\hbox{\lower4pt\hbox{$\sim$}}}\hbox{$>$}}}}
\newbox\grsign \setbox\grsign=\hbox{$>$} \newdimen\grdimen
\grdimen=\ht\grsign
\newbox\simlessbox \newbox\simgreatbox \newbox\simpropbox
\setbox\simgreatbox=\hbox{\raise.5ex\hbox{$>$}\llap
     {\lower.5ex\hbox{$\sim$}}}\ht1=\grdimen\dp1=0pt
\setbox\simlessbox=\hbox{\raise.5ex\hbox{$<$}\llap
     {\lower.5ex\hbox{$\sim$}}}\ht2=\grdimen\dp2=0pt
\setbox\simpropbox=\hbox{\raise.5ex\hbox{$\propto$}\llap
     {\lower.5ex\hbox{$\sim$}}}\ht2=\grdimen\dp2=0pt

\title[Multiwaveband photometry of WD0137-349B ]{Multiwaveband photometry of the  irradiated brown dwarf  WD0137-349B}
\author[S. L. Casewell et al.]{S. L. Casewell$^{1}$  \thanks{E-mail:
slc25@le.ac.uk},  K.A. Lawrie $^{1}$,  P. F. L. Maxted $^{2}$, M. S. Marley$^{3}$, J. J.  Fortney$^{4}$, P. B. Rimmer $^{5}$, 
\newauthor  S. P. Littlefair $^{6}$, G. Wynn $^{1}$, M. R. Burleigh $^{1}$ and Ch. Helling $^{5}$\\
$^{1}$Department of Physics and Astronomy, University of Leicester, University Road, Leicester LE1 7RH, UK\\
$^{2}$Department of Physics and Astrophysics, Keele University, Keele, Staffordshire, ST5 5BG, UK\\
$^{3}$NASA Ames Research Center, MS-245-3, Moffett Field, CA 94035, USA \\
$^{4}$Department of Astronomy and Astrophysics, University of California, Santa Cruz, CA 95064, USA\\
$^{5}$SUPA, School of Physics and Astronomy, University of St Andrews, St Andrews, KY16 9SS, UK\\
$^{6}$Department of Physics and Astronomy, University of Sheffield, Sheffield, S3 7RH, UK\\
}

\begin{document}

\date{\today}

\pagerange{\pageref{firstpage}--\pageref{lastpage}} \pubyear{2014}

\maketitle

\label{firstpage}

\begin{abstract}

WD0137-349 is a white dwarf-brown dwarf binary system in a 116 minute orbit. We present radial velocity observations and multiwaveband photometry from $V$, $R$ and $I$ in the optical, to $J$, $H$ and $K_s$ in the near-IR and [3.6], [4.5], [5.8] and [8.0]
$\mu$m in the mid-IR.  The photometry and lightcurves show variability in all wavebands, with the amplitude peaking at [4.5] microns, where the system is also brightest. Fluxes and brightness temperatures were computed for the heated and unheated atmosphere of the brown dwarf (WD0137-349B) using synthetic spectra of the white dwarf using model atmosphere simulations.  We show that the flux from the brown dwarf dayside is  brighter than expected in the $K_s$ and [4.5] $\mu$m  bands when compared to models of irradiated brown dwarfs with full energy circulation and suggest this over-luminosity  may be attributed to  H$_{2}$ fluorescence or H$_{3}^{+}$ being generated in the atmosphere by the UV irradiation.

\end{abstract}


\section{Introduction}

WD0137-349 is one of only 4 known detached post-common envelope binaries containing a white dwarf and a brown dwarf companion. The other three are: GD1400, (WD+L6, P=9.98hrs \citealt{farihi,  dobbie, burleigh11}), WD0837+185 (WD+T8, P=4.2hrs \citealt{casewell12}) and NLTT5306 (WD+L4-L7, P=101.88 min \citealt{steele13}). Another eclipsing, candidate system has recently been reported (CSS21055, \citealt{beuermann14}) but is yet to be confirmed via radial velocity or spectroscopy. 

WD0137-349 was discovered in 2006 by \citet{maxted06} and better characterised by \citet{burleigh06} using spectra from the Gemini telescope and the Gemini Near-InfraRed Spectrograph (GNIRS).  The system is composed of a 0.4M$_{\odot}$ DA white dwarf (T$_{\rm eff}$=16500 K) and a secondary of mass 53 M$_{\rm Jup}$. The period of the system is 116 minutes and the estimated separation is 0.65R$_{\odot}$ \citep{maxted06}. \citet{burleigh06} determined from the GNIRS spectra that an L8 is the most likely spectral type for the brown dwarf, and that it has not likely been affected by common envelope evolution. Such systems are not only important from an evolutionary point of view, giving us insight into the common envelope parameters (e.g. \citealt{casewell12}), but also provide insight into the properties of irritated atmospheres.
WD0137-349B is a brown dwarf that intercepts $\sim$1 per cent of the light from its white dwarf companion. Due to the high effective temperature (16500 K) of the white dwarf, this light is primarily in the ultraviolet regime. As the system is likely tidally locked, this means one side of the brown dwarf is continually irradiated by the white dwarf.  This situation is similar to that in many exoplanets such as Wasp-12 b \citep{hebb09} and HD189733b \citep{knutson12}.	
Observing white dwarf+brown dwarf binaries have two advantages over similar exoplanet systems. The first is that  brown dwarf atmospheres are well characterised  (e.g. \citealt{cushing04}), and that brown dwarf atmosphere modelling is more advanced than that of exoplanets.  The second is that brown dwarfs are relatively bright so we can directly observe the brown dwarf in the $K$ band and at longer wavelengths where the brown dwarf is brighter than its white dwarf companion (\citealt{maxted06, burleigh06, casewell13}). 

\section{Observations and  Data Reduction }
\subsection{UVES spectra}
 We obtained 50 spectra of WD0137-349 through service mode observations made
with the {\sc uves} echelle spectrograph \citep{dekker00} mounted
on UT2 (`Kueyen') of  the Very Large Telescope (VLT) at the European Southern
Observatory, Paranal, Chile (ESO program ID 079.C-0683). We used a 1.4\,arcsec
wide slit for these observations so the resolution of the spectra is set by
the seeing. The ambient seeing during the exposures varied from 0.7 to 3.5
arcsec with a median value of 1.3 arcsec, corresponding to a resolution of
9.5\kms. We used the CD4 cross-dispersing grating to observe the wavelength
range  5655\,--\,9460\AA\ on the red arm. We do not use the blue arm data in
this analysis.  The exposure time was 240 seconds per spectrum. Spectra were
reduced using the {\sc  uves} pipeline version 4.3.0 provided by the
observatory. The dispersion near the H$\alpha$ line is 0.072\AA-per-pixel and 
typical signal-to-noise per pixel is about 20.

 We also re-analysed the 20 UVES spectra reduced automatically at the
observatory described in \citet{maxted06} (ESO program ID
276.D-5014). These spectra have an exposure time of 290 seconds and were
obtained with the CD2 cross-dispersing grating, but are otherwise similar to
the spectra described above.

\begin{figure}
\mbox{\includegraphics[width=0.49\textwidth]{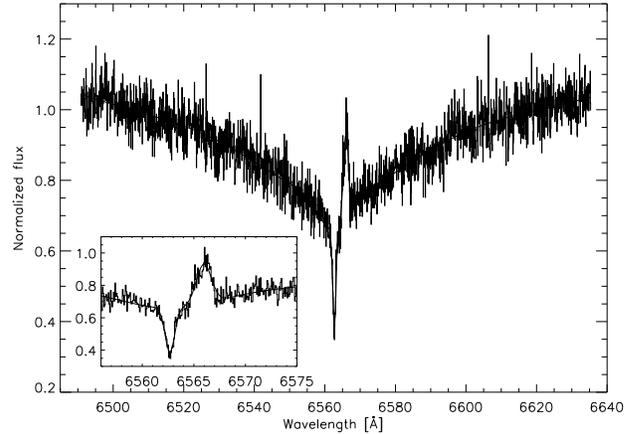}}
\caption{
  Least-squares fit to the H$\alpha$ line for a typical UVES spectrum of
WD0137$-$349. The observed spectrum is plotted as a histogram and the
least-squares fit as a smooth line. Inset: detail showing the narrow core of
the white dwarf absorption line and the emision line from the brown dwarf. 
\label{fitspec}}
\end{figure}

 We used the program {\sc molly}\footnote{\it
deneb.astro.warwick.ac.uk/phsaap/software} to measure the radial velocities of
the stars by fitting multiple Gaussian profiles to the H$\alpha$ line. We used
three Gaussian profiles with the same mean but independent widths to model the
white dwarf absorption line and a single Gaussian profile for the emission
line from the brown dwarf. Free parameters in the fit were the coefficients of
a low-order polynomial used to model the continuum level, the depths of the
two widest Gaussian profiles, the height of the Gaussian profile used to model
the emission line from the brown dwarf and the positions of the emission and
absorption lines. Other parameters for the Gaussian profiles were fixed at
values determined from an preliminary least-squares fit. The
least-squares fit to a typical spectrum is shown in Fig.~\ref{fitspec}.
The radial velocities for the white dwarf and the emission line from the 
brown dwarf derived from these least-squares fits are given in
Table~\ref{rvtable}.

\begin{figure}
\mbox{\includegraphics[width=0.49\textwidth]{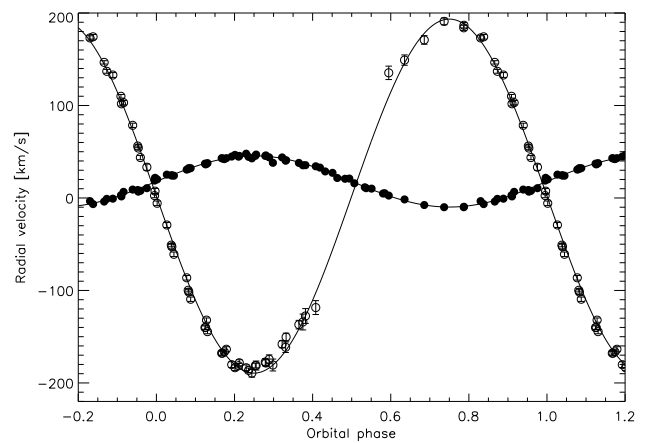}}
\caption{
  Least-squares fits of sine functions to the measured radial velocities of
the white dwarf (filled circles) and the emission line from the brown dwarf
(open circles).
\label{rvfitfig}}
\end{figure}

We fitted a sine function of the form $V_r({\rm WD}) = \gamma + K_{\rm
WD}\sin(2\pi(t-T_0)/P)$ to the measured radial velocities of the white dwarf
to determine the orbital period, $P$, and a reference time when the brown dwarf
is at the furthest point in its orbit from the observer, $T_0$. We added
1\,km/s in quadrature to the standard errors given in Table~\ref{rvtable} in
order to achieve a reduced chi-squared value for the fit $\chi^2_r = 1$. The
parameters of this fit are given in Table~\ref{rvfittable}. Also given in this
table is a fit to the radial velocities measured for the emission line from
the brown dwarf with the values of $T_0$ and $P$ fixed to the values
determined from the fit to the measured radial velocities of the white dwarf.
We excluded measurements with large standard errors from this fit -- these
measurements come from spectra obtained at orbital phases where the emission
line is very weak.
To obtain $\chi^2_r = 1$ for the least-squares  fit to the emission line
radial velocities we had to add 5.5\kms\ in quadrature to the standard errors
given in Table~\ref{rvtable}. This much larger ``systematic error'' is mostly
the result of the variations in the shape of the line profile during the
orbit that are not accounted for by fitting the line with a Gaussian profile.
The fit to both sets of radial velocities are shown in Fig.~\ref{rvfitfig}. 

\begin{table*} \caption{Radial velocity measurements for the white dwarf
($V_r({\rm WD})$) and the emission line from the brown dwarf ($V_r({\rm em})$)
derived from multiple Gaussian profiles fit by least-squares to the H$\alpha$
line. \label{rvtable} }
\begin{tabular}{@{}rrrcrrr}
\hline
 \multicolumn{1}{l}{HJD}
&\multicolumn{1}{l}{$V_r(\rm WD)$}
&\multicolumn{1}{l}{$V_r(\rm em)$} 
&~~~&\multicolumn{1}{l}{HJD}
&\multicolumn{1}{l}{$V_r(\rm WD)$}
&\multicolumn{1}{l}{$V_r(\rm em)$} \\
\multicolumn{1}{l}{$-2450000$}
&\multicolumn{1}{c}{[km\,s$^{-1}$]}
&\multicolumn{1}{c}{[km\,s$^{-1}$]} 
&&\multicolumn{1}{l}{$-2450000$}
&\multicolumn{1}{c}{[km\,s$^{-1}$]}
&\multicolumn{1}{c}{[km\,s$^{-1}$]} \\
\noalign{\smallskip}\hline
3686.5258&$ 10.6\pm 1.8$&$  33.3\pm3.4$&&4366.7203&$ 32.8\pm 1.5$&$ -79.1\pm 10.2 $\\
3686.5299&$ 25.2\pm 1.5$&$ -29.2\pm2.7$&&4366.7238&$ 21.6\pm 1.7$&$  70.9\pm 21.6 $\\
3686.5339&$ 30.7\pm 1.3$&$ -86.3\pm2.5$&&4366.7273&$ 15.9\pm 1.9$&$  28.4\pm 44.5 $\\
3686.5379&$ 36.9\pm 1.2$&$-132.1\pm2.3$&&4366.7308&$ 10.1\pm 1.6$&$  36.2\pm 24.6 $\\
3686.5420&$ 43.0\pm 1.2$&$-163.6\pm2.7$&&4366.7343&$  2.6\pm 1.4$&$ 135.3\pm  7.3 $\\
3686.5460&$ 47.9\pm 1.2$&$-183.6\pm3.1$&&4378.7497&$ -3.9\pm 1.0$&$ 146.6\pm  1.8 $\\
3686.5500&$ 46.0\pm 1.2$&$-178.4\pm3.5$&&4378.7532&$  2.9\pm 1.0$&$ 110.0\pm  2.0 $\\
3686.5540&$ 40.3\pm 1.2$&$-161.4\pm5.7$&&4378.7566&$  8.6\pm 1.2$&$  56.2\pm  1.9 $\\
3686.5580&$ 35.5\pm 1.2$&$-127.6\pm8.0$&&4378.7600&$ 18.8\pm 1.5$&$   2.6\pm  2.3 $\\
3686.5621&$ 28.7\pm 1.2$&$ -84.2\pm0.6$&&4378.7634&$ 24.4\pm 1.1$&$ -51.0\pm  1.8 $\\
3686.5661&$ 20.7\pm 1.4$&$ -10.3\pm8.8$&&4378.7668&$ 31.2\pm 1.0$&$ -99.8\pm  1.8 $\\
3686.5701&$ 11.6\pm 1.4$&$  90.1\pm9.4$&&4378.7702&$ 37.1\pm 0.9$&$-140.2\pm  1.8 $\\
3686.5741&$  5.1\pm 1.2$&$ 121.0\pm1.2$&&4378.7736&$ 43.0\pm 0.9$&$-167.8\pm  2.0 $\\
3686.5782&$ -1.6\pm 1.2$&$ 149.1\pm5.6$&&4378.7771&$ 44.6\pm 1.0$&$-181.4\pm  2.4 $\\
3686.5822&$ -7.6\pm 1.1$&$ 170.9\pm4.6$&&4378.7805&$ 45.6\pm 1.0$&$-181.2\pm  2.5 $\\
3686.5862&$-10.0\pm 1.1$&$ 190.9\pm3.6$&&4387.6494&$  1.7\pm 1.2$&$ 101.8\pm  2.3 $\\
3686.5903&$ -9.5\pm 1.1$&$ 186.8\pm2.9$&&4387.6528&$  6.8\pm 1.4$&$  54.0\pm  2.1 $\\
3686.5943&$ -6.6\pm 1.1$&$ 174.3\pm2.5$&&4387.6563&$ 21.8\pm 2.0$&$   7.4\pm  2.9 $\\
3686.5983&$ -0.8\pm 1.2$&$ 133.0\pm2.6$&&4387.6597&$ 24.5\pm 1.5$&$ -52.7\pm  2.7 $\\
3686.6023&$  9.1\pm 1.4$&$  78.6\pm2.3$&&4387.6631&$ 32.3\pm 1.3$&$-101.7\pm  2.5 $\\
4343.8265&$ 44.8\pm 1.2$&$-180.2\pm3.1$&&4387.6665&$ 36.3\pm 1.4$&$-140.1\pm  2.6 $\\
4343.8299&$ 44.5\pm 1.2$&$-186.4\pm2.5$&&4387.6700&$ 42.8\pm 1.3$&$-167.8\pm  2.7 $\\
4343.8333&$ 45.5\pm 1.1$&$-177.4\pm3.6$&&4387.6734&$ 45.5\pm 1.3$&$-178.2\pm  3.3 $\\
4343.8367&$ 43.9\pm 1.1$&$-158.1\pm3.8$&&4387.6768&$ 47.0\pm 1.3$&$-181.2\pm  4.9 $\\
4343.8401&$ 37.9\pm 1.1$&$-137.0\pm4.6$&&4387.6802&$ 38.0\pm 1.8$&$-180.6\pm  6.4 $\\
4343.8436&$ 34.3\pm 1.1$&$-118.5\pm7.5$&&4400.5867&$-10.0\pm 1.1$&$ 184.0\pm  2.8 $\\
4343.8470&$ 27.2\pm 1.2$&$ -56.1\pm4.6$&&4400.5901&$ -3.4\pm 1.1$&$ 173.1\pm  2.4 $\\
4343.8504&$ 21.0\pm 1.2$&$ -63.2\pm3.2$&&4400.5935&$ -1.0\pm 1.1$&$ 136.9\pm  2.1 $\\
4343.8538&$ 10.6\pm 1.5$&$  40.3\pm2.4$&&4400.5969&$  6.6\pm 1.2$&$ 103.1\pm  2.4 $\\
4343.8572&$  4.8\pm 1.4$&$  86.7\pm1.5$&&4400.6004&$  7.7\pm 1.5$&$  43.7\pm  2.3 $\\
4366.7030&$ 46.6\pm 1.5$&$-183.5\pm2.9$&&4400.6038&$ 20.1\pm 1.6$&$  -5.7\pm  2.5 $\\
4366.7064&$ 42.8\pm 1.6$&$-189.4\pm4.4$&&4400.6072&$ 24.1\pm 1.3$&$ -60.9\pm  2.5 $\\
4366.7099&$ 43.9\pm 1.3$&$-174.4\pm4.7$&&4400.6106&$ 32.2\pm 1.1$&$-109.5\pm  2.4 $\\
4366.7134&$ 40.8\pm 1.4$&$-150.3\pm4.2$&&4400.6140&$ 37.7\pm 1.1$&$-144.5\pm  2.3 $\\
4366.7168&$ 35.3\pm 1.4$&$-134.1\pm8.6$&&4400.6174&$ 42.0\pm 1.1$&$-166.4\pm  2.1 $\\
\hline
\hline
\end{tabular}
\end{table*}

\begin{table}
 \caption{Parameters for least-squares fits of a circular orbit to our
measured radial velocities, $V_r = \gamma + K\sin(2\pi (t-T_0)/P)$. Values in
parentheses are standard errors on the final digit. 
\label{rvfittable}
}
 \begin{tabular}{@{}lrr}
\hline
Parameter & \multicolumn{1}{l}{White dwarf} &
\multicolumn{1}{l}{Brown dwarf} \\
\hline
$T_0$ (HJD)  & 2454178.6762(1) & -- \\
$P$ [days]   & 0.07943002(3) & -- \\
$\gamma$ [\kms] & 17.9(2) & 3(1) \\
$K$ [\kms]      &  27.7(3) & $-192(2)$ \\
\hline
\end{tabular}
\end{table}

\subsection{The optical imaging and its reduction}
The optical data was observed using the South African Astronomical Observatory (SAAO) 1.0m telescope and the SAAO CCD (STE3) instrument which has a CCD size of 512 $\times$ 512 pixels and a pixel scale of 0.31 arcsec/pixel.  The observations are detailed in Table \ref{optical}.

\begin{table}
\caption{\label{optical} Observations from the SAAO.}
\begin{center}
\begin{tabular}{c c c c c}
\hline
Date & Filter &  Number of points& Exposure time (s) & S/N\\
\hline
2007-11-28  &   $I$	&29    & 200	&100\\	
2007-12-04   &  $V$	&79    & 90	&104\\	
2008-11-08   &  $I$	&322   & 42	&70	\\
2008-11-09   & $R$	&238   & 42	&118\\	
2008-11-10   &  $I$&137    &42	&79	\\
\hline
\end{tabular}
\end{center}
\end{table}

The data were reduced using the SAAO CCD pipeline which subtracted the bias and normalised the science frames with the master flat field frame. We used the \textsc{starlink} package \textsc{autophotom} to perform the photometry of the target and comparison stars. The aperture was fixed for the data and was set to be 2 times the mean seeing (full width, half maximum: \citealt{naylor98}). This aperture size limited the impact of the background noise in the aperture. The sky background level was determined using the clipped mean of the pixel value in an annulus around the stars and the measurement errors were estimated from the sky variance. To remove atmospheric fluctuations, the light curve was divided by the light curve of one of the comparison stars. 

\subsection{The near-IR imaging and its reduction}

Photometry was obtained using Son OF Isaac (SOFI: \citealt{moorwood98}) on the New Technology Telescope at La Silla on the nights of  2007-10-02 to  2007-10-04 and 2007-10-24 to 2007-10-26 .  The seeing was typically 1.5"-2.0", and the data were obtained in the $J$, $H$ and $K_s$ bands. The exposure times were 2s in $J$, 4s in $H$ and 8s in $K_{s}$, in a 5 point jitter pattern which was then combined into mosaics to remove the sky background. The data were reduced using the \textsc{starlink} package \textsc{orac\_dr} with the SOFI specific routines to perform the flat fielding, sky subtraction and mosaic combining. Object extraction was performed using aperture photometry routines within SExtractor and an aperture equivalent to the seeing. 

To combine the $K_{s}$ data from each night each frame was calibrated using stars within the field from the 2 Micron All Sky Survey \citep{cutri03}.  The $J$ and $H$ lightcurves are all from the same night. There are 130 points in the $K_s$ band lightcurve, and 14 in the $J$ and $H$ lightcurves.

\subsection{The Spitzer imaging and its reduction}
We have also obtained $Spitzer$  IRAC \citep{fazio04} photometry  from Cycle 7
(Programme ID:40325,  PI: Burleigh) in all four bands  ([3.6], [4.5], [5.8] and
[8.0] microns). Each waveband was observed for one 116 minute orbit.  The data were observed using 30s integrations, the full array and no
dithering, as time series photometry was required.

For the [3.6] and [4.5] micron data no additional processing was required and
aperture photometry was performed on each individual image using the
\textsc{apex} software and an aperture of 3 pixels with a background aperture
of 12-20 pixels. Pixel phase, array location dependence, and aperture
corrections were applied to the data and the IRAC zero magnitude flux
densities as found on the $Spitzer$ website were then used to convert the flux
into magnitudes on the Vega magnitude scale. The photometric errors were
estimated using the Poisson noise given by the  \textsc{apex} software which
was then combined with the errors on the zeropoints. The 3 per cent absolute
calibration was added in quadrature to these photometric errors. The S/N of
these data is typically 40.

For the [5.8] and [8.0] micron bands, the S/N was not high enough to perform
aperture photometry on the individual images so the recommended \textsc{mopex} pipeline
(v18.3.6 final)  was used to perform overlap corrections and to create final
and array correction mosaics of the images, combining 20 images at a time resulting in a total exposure time of 600s and S/N of 10-20. Aperture photometry was performed
as for the [3.6] and [4.5] micron images. 
In total, we have 240 data points in the [3.6] and [4.5] micron bands, and 12 in
the [5.8]  and [8.0] micron bands.

\section{Results}
We converted all of the UTC Julian dates in the image headers to BJD and  fitted the $V$, $R$ and $I$ lightcurves to the ephemeris derived from the radial velocity measurements of the hydrogen absorption lines in the white dwarf.  These are seen in Figure \ref{optical}.
We fitted each night of data separately. The variability was determined by fitting a sine curve to the data. The peak-to-peak amplitude was measured to be 1.22$\pm$0.3 per cent in $V$, 3.39$\pm$0.23 per cent in $I$ and 2.11$\pm$0.15 per cent in $R$ (Figure \ref{optical}).  The light curve in the $I$ band from 10-11-2008 has a peak-to-peak amplitude of  3.31$\pm$0.33 per cent which is consistent with the data from 08-11-2008. The $I$ band light curve from 28-11-07 has a variability of 4.05$\pm$0.51 per cent which is slightly larger than the other two nights of $I$ band data, but is still just within the errors. This lightcurve only has very few points (29) which do not cover a whole orbit of the system and a longer exposure time which means the cadence of the data is not as good as for the other $I$ band data.

\begin{figure}
\begin{center}
\scalebox{0.35}{\includegraphics[angle=0, scale=1.4]{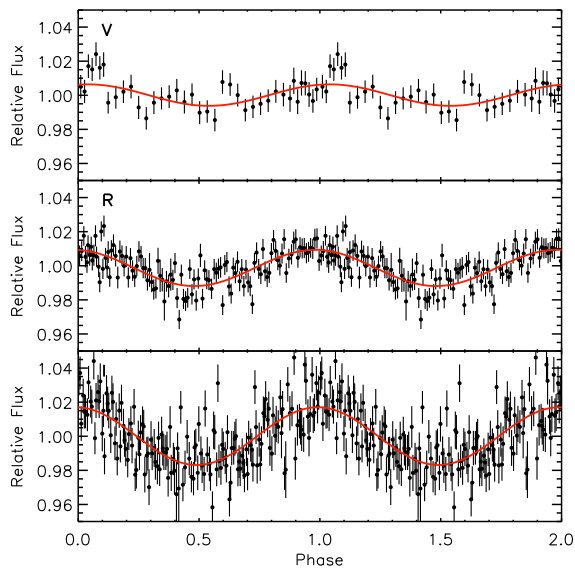}}
\caption{\label{optical}Two cycles of the $V$, $R$ and $I$ lightcurves of WD0137-349 folded on the ephemeris of the system (solid line), and binned by a factor of two.  Only the $I$ band data from 2008-11-08 are shown here. }
\end{center}
\end{figure}

\begin{figure}
\begin{center}
\scalebox{0.35}{\includegraphics[angle=270]{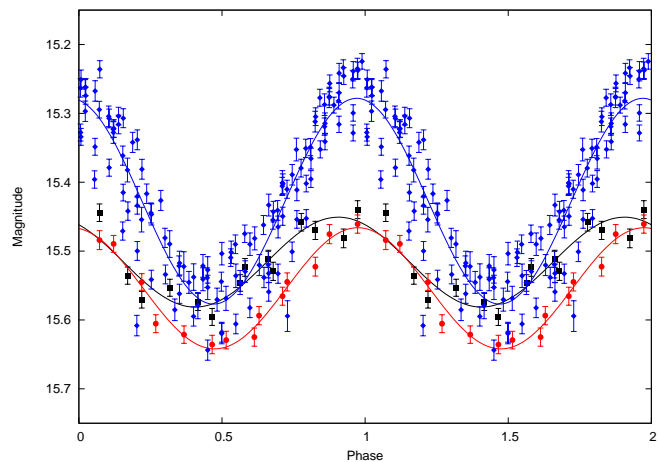}}
\caption{\label {nir} $J$ (black squares), $H$ (red circles) and $K_{s}$ (blue diamonds) lightcurves of WD0137-349 folded on the ephemeris of the system. }
\end{center}
\end{figure}

\begin{figure}
\begin{center}
\scalebox{0.35}{\includegraphics[angle=270]{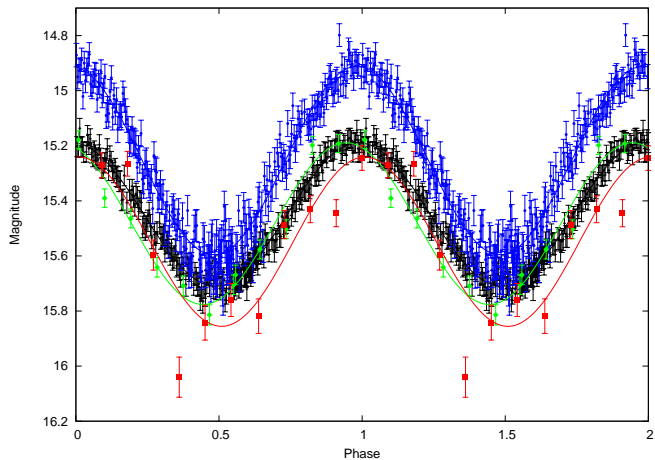}}
\caption{\label {spitzer} [3.6] (black +), [4.5] (blue circles), [5.8] (green diamonds) and [8.0] (red squares) micron lightcurves of WD0137-349 folded on the ephemeris of the system. }
\end{center}
\end{figure}

We fitted all of the near-infrared (Figure \ref{nir}) and mid-infrared (Figure \ref{spitzer}) data as above.   All the photometry is in phase, and shows no statistically significant lag as is seen in  Table \ref{lag}. These data show that the irradiation is inducing a day/night contrast in the atmosphere of the brown dwarf. This is supported by the H$\alpha$ detection.

\begin{table}
\caption{\label{lag} Phase difference of light curve fit compared to the ephemeris derived from the radial velocity.}
\begin{center}
\begin{tabular}{c c c}
\hline
Filter & semi-amplitude&$\phi$\\
\hline
$J$&0.06&0.064$\pm$0.020\\
$H$&0.08&0.016$\pm$0.009\\
$K_s$& 0.15& 0.004$\pm$0.008\\
{[3.6]}& 0.26& 0.003$\pm$0.002\\
{[4.5]}&  0.34& 0.009$\pm$0.002\\
{[5.8]}&  0.29 &0.046$\pm$0.010\\
{[8.0]}&  0.31& 0.076$\pm$0.087\\
\hline
\end{tabular}
\end{center}
\end{table}

To determine the brown dwarf flux we generated a pure hydrogen white dwarf model using the plane-parallel, hydrostatic, 
non-local thermodynamic equilibrium atmosphere and spectral synthesis codes \textsc{tlusty} \citep{hubeny98, hubeny95} and \textsc{synspec} \citep{hubeny01} for an effective temperature of 16500 K and log g of 7.49 \citep{maxted06}. We then convolved this spectum with the filter profiles from 2MASS and IRAC, to get the total white dwarf flux per filter. These flux measurements, including a 12 per cent  error to take into account the error in the effective temperature (500 K) and log g (0.08) as in \citet{maxted06},  were then compared to the theoretical colour tables for white dwarfs  \citep{bergeron, kowalski06, tremblay11} which have been extended into the $Spitzer$ bandpasses (P. Bergeron, private comm) and was also found to be consistent. 
The total white dwarf flux per filter was then subtracted from the observed flux (white dwarf + brown dwarf combined) at the maximum and minimum values. This flux value was then converted back into magnitudes (Table \ref{mags2}).  

\begin{table*}
\caption{\label{mags2}Magnitudes and brightness temperatures of the combined white dwarf+brown dwarf system, the white dwarf magnitudes determined from the model and the irradiated and un-irradiated hemispheres of WD0137-349B after the white dwarf has been subtracted. The errors on the brightness temperatures and flux reflect the error in the photometry and the uncertainty on the white dwarf temperature and gravity. We were unable to measure the nightside temperature in the $H$ and [5.8] $\mu$m filters, and so an upper limit of the dayside flux has been used.}
\begin{center}
\begin{tabular}{c c c c c c c c}
\hline
Waveband &\multicolumn{2}{c}{Magnitude (WD+BD) }& Magnitude (WD)& \multicolumn{2}{c}{Magnitude (BD)}&\multicolumn{2}{c}{Brightness Temperature (K)}\\
& Dayside& Nightside&&Dayside & Nightside & Dayside & Nightside\\ 

\hline
$J$&15.45$\pm$0.01&15.59$\pm$0.01&15.72$\pm$0.12&17.14$^{+0.42}_{-0.84}$&17.97$^{+0.87}_{-3.39}$&2418$_{-329}^{+201}$&2085$_{-769}^{+287}$\\
$H$&15.47$\pm$0.01&15.65$\pm$0.01&15.52$\pm$0.12&18.33$\pm$0.80& 18.33&1585$\pm$329&1585\\
$K_{s}$&15.28$\pm$0.01&15.57$\pm$0.01&15.77$\pm$0.12&16.37$\pm$0.29&17.49$^{+0.56}_{-1.16}$&2015$_{-131}^{+119}$&1537$_{-262}^{+183}$\\
{[3.6]}&15.19$\pm$0.04&15.71$\pm$0.04&15.82$\pm0.12$&16.03$_{-0.17}^{+0.34}$& 18.04$_{-0.82}^{+1.49}$&1668$_{-110}^{+103}$& 958$_{-508}^{+184}$\\
{[4.5]}&14.91$\pm$0.04&15.60$\pm$0.06&15.85$\pm$0.12&15.50$_{-0.15}^{+0.19}$&17.31$_{-0.61}^{+1.10}$&1808$_{-93}^{+91}$&974$_{-180}^{+138}$\\
{[5.8]}&15.19$\pm$0.03&15.78$\pm$0.04&15.76$\pm$0.12&16.12$_{-0.16}^{+0.43}$&16.12&1254$_{-101}^{+97}$&1254\\
{[8.0]}& 15.85$\pm$0.06&15.24$\pm$0.05&15.82$\pm$0.12&16.06$_{-0.1}^{+0.59}$&19.88$\pm$2.2&1102$_{-102}^{+99}$&364$\pm$249\\

\hline
\end{tabular}
\end{center}
\end{table*}

It is evident from this calculation that we do not detect any flux from the nightside side of the brown dwarf in the $H$ or [5.8] micron wavebands, and the values for the dayside of the brown dwarf have been used as upper limits. This is based on the assumption that the nightside cannot be brighter than the dayside.  Using the magnitudes from Table \ref{mags2} we calculated the brightness temperature at each wavelength from the $J$ band to [8.0] microns on the heated and unheated side of WD0137-349B and they can also be seen in Table \ref{mags2} and Figure \ref{model}. These temperatures seem to imply that there is a weak dayside temperature inversion between the scale heights probed in the $K_{s}$ and [4.5] micron bands.  
This is not uncommon in hot Jupiter atmospheres and is normally attributed to the presence of TiO/VO opacity \citep{fortney08, hubeny03}.

In \citet{burleigh06} we estimated the equilibrium temperature of the irradiated side of WD0137-349B.  The equilibrium temperature is the temperature of the heated side if the white dwarf and brown dwarf are perfect black bodies and is calculated to be $\sim$2000 K. This is too hot for an L6 dwarf, but this estimate is broadly consistent with our brightness temperature calculations.  While Figure 2 of \cite{casewell13} shows that $JHK$ photometry of the  heated face of the brown dwarf is consistent with a L6 dwarf, it is likely that there are other mechanisms occurring within the brown dwarf atmosphere, particularly considering that while one hemisphere is being irradiated, energy  is being  redistributed around the object as has been suggested for some exoplanets (e.g. \citealt{cooper05}).

\section{Modelling the data}

The likely atmospheric circulation and irradiation of WD0137-349B make it unreasonable to compare the photometry of WD0137-349B with that of unirradiated brown dwarfs, as had been done in the past \citep{burleigh06}.  Therefore we have used models that take into account the high level of UV irradiation in this system and use them to explore the heat transport and potential temperature inversion as well as the effects of some photochemistry.

\subsection{The irradiated hemisphere}

\begin{figure}
\begin{center}
\scalebox{0.35}{\includegraphics[angle=270]{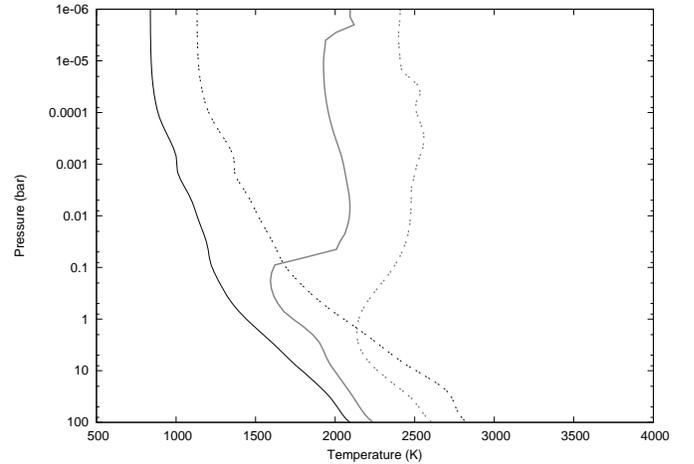}}
\caption{\label {model2} The temperature-pressure profiles for the dayside of the irradiated brown dwarf model.  Solid lines are models that use  full circulation and energy transport from the heated to non-heated side of the brown dwarf. Dotted lines show the zero circulation models. The grey lines are models containing TiO and black lines for the non-TiO model.}
\end{center}
\end{figure}

\begin{figure}
\begin{center}
\scalebox{0.35}{\includegraphics[angle=270]{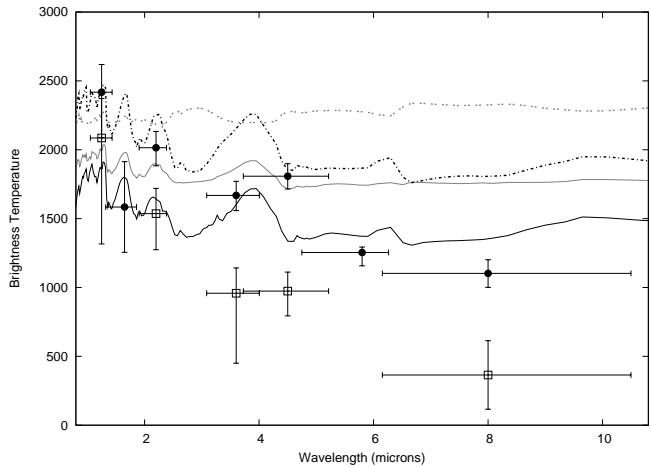}}
\caption{\label {model} Brightness temperatures for WD0137-349 on the irradiated (circles) and unirradiated (open boxes) sides of the brown dwarf. The $H$ and [5.8] temperatures for the unirradiated side are upper limits only (diamonds), derived from the white dwarf's flux.  The errorbars on the X scale represent the widths of the filters used. The models are as for Figure \ref{model2} and show the dayside only. }
\end{center}
\end{figure}

To explore the response of a brown dwarf atmosphere to the intense irradiation environment and to explore the possibility of a 
temperature inversion we computed a few exploratory models using the atmospheric structure model of \citet{marley99}, \citet{marley02} 
and \citet{fortney05} with a log g=5.32 surface gravity, and intrinsic effective temperatures ranging from 600 to 1000 K.  The choice of 
intrinsic effective temperature does not affect the results as in all models the incident flux dominates the energy budget and is far larger 
than the intrinsic flux.  The white dwarf parent star has a T$_{\rm eff}=$16,500 K, placing the peak of its Planck function well into the UV.  
The model atmosphere code we use for the brown dwarf, which was generally developed for Sun-like stars, lacks opacities at such short wavelengths.  To model 
the incident flux we instead used a cooler 10,000 K blackbody with a larger white dwarf radius such that same total incident flux, but 
peaking at a longer wavelength, was maintained for the brown dwarf at 0.003 AU. The temperature-pressure profiles are shown in 
Figure \ref{model2}.  As is common for hot Jupiters (e.g., \citealt{fortney06}) we explored two cases for re-radiation of the absorbed flux 
on the irradiated atmosphere.  In some models, the incident flux is cut by one-half, to simulate energy being efficiently lost to the 
un-modelled night side (full circulation models).  In others, the full incident flux is used, to simulate no energy lost to the night side, 
which leads to a hotter day side model. Our aim was not to reproduce the observed photometry but rather to understand the type of 
response that would be expected in radiative-convective and thermochemical equilibrium.  As is also customary for irradiated hot 
Jupiter models we considered cases in which the brown dwarf's atmosphere both contained and lacked gaseous TiO. Since TiO 
absorbs strongly at optical wavelengths, it has been suggested as the absorber responsible for creating atmospheric thermal inversions 
in some hot Jupiters \citep{fortney08, hubeny03}. Model brightness temperature spectra for these cases are shown in Figure \ref{model}.

It is clear that the models with the full circulation, where energy is being lost to the night side, produce brightness temperatures closer to those observed. The data also show a larger dayside brightness temperature for $K_{s}$ band than for [4.5] $\mu$m. Since the opacity is larger in the former bandpass than the latter, this implies a temperature profile that increases with altitude above a minimum. 
The difference is relatively small, $\sim$200 K, but is similar to that seen in exoplanets (e.g. \citealt{madu11}). Surprisingly, however,
the overall flux levels are better matched by the models lacking in TiO, which raises a question as to which absorber would
then be responsible for the inversion. Furthermore the condensation temperature of TiO is $\sim$1670 K, therefore in WD0137-349B, TiO (and the similar VO) should be present in the atmosphere.

Two explanations for this unexpected behaviour seem likely. First, it is possible that when WD0137-349B orbited the white dwarf progenitor that the TiO and VO condensed out of the upper atmosphere. The night side photometry indicates that  it may have had an effective temperature of 1000 K, cool enough for condensation to occur. 
It is also possible that the temperature inversion is not real, and  that the greater brightness temperature
at K$_{s}$ is in fact due to over luminosity   caused by photochemistry  which is not included in these models.  

The $J$ band also appears anomalously high, but it can be seen from the spectra in \citet{casewell13, burleigh06} that the brown dwarf is contributing very little of the flux at these wavelengths, and what we are likely detecting is scattered light off the  brown dwarf atmosphere. Despite the majority of the incident light from the white dwarf being emitted in the UV, it is unlikely that we will see
some sort of photochemical haze as predicted by \citet{zahnle} as these should dissipate at effective temperatures greater than 1000 K. 
Is is possible, however that 
we are seeing some form of airglow such as the H$_{3}^{+}$ seen at Jupiter \citep{drossart89}, Saturn \citep{trafton93} and Uranus \citep{geballe93}
which emits at wavelengths between 2 and 6 microns, and is strongest between 3 and 5 microns (\citealt{Neale1996}, calculated as part of the exomol project \citealt{tennyson12}) at the temperature and pressure of a mid-L dwarf.  Another possibility for the over-luminosity is that we see H$_{2}$ fluorescence as was recently detected in the UV  active M dwarf exoplanet hosts \citep{france13}. However, it has not been determined whether the fluorescence is due to the M dwarf activity, or the irradiation of the planet.

\subsection{Photochemistry}
\begin{figure}

\begin{center}

\scalebox{0.55}{\includegraphics{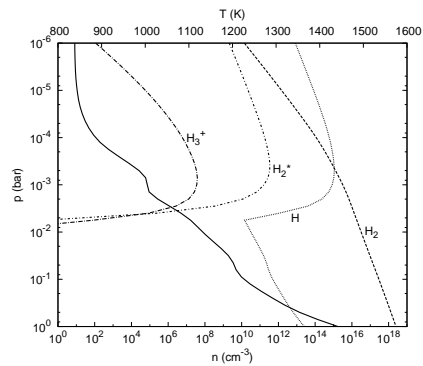}}

\caption{\label {h2-abundance} The number densities [cm$^{-3}$] of H, H$_2$, and H$_3^+$
and UV-pumped H$_2$ (denoted by H$_2^*$) (bottom x-axis) as a function of pressure [bar] (y-axis), as 
calculated by applying the chemical network of \citet{Rimmer2014} to the (non-TiO) temperature-pressure 
profile of \ref{model2}. The temperature-pressure profile used is shown here as a sold line 
(gas temperature [K] in the top x-axis).}

\end{center}

\end{figure}

We utilise the temperature-pressure profile for the dayside of the irradiated brown 
dwarf model without TiO, using full convection and energy transport (the solid black line in 
Figure \ref{model2}). The ion-neutral chemistry of \citet{Rimmer2014} is applied on top of this 
temperature-pressure profile non-self-consistently, with the photochemistry from 
\citet[][and references within]{yelle04}. Rate coefficients for photolysis due to the the 
irradiating white dwarf are calculated by assuming that the white dwarf produces a UV field as a 
black body with $T_{\rm irr} = 16500$ K. The UV field is attenuated by H$_2$ self-shielding, using 
shielding coefficients given by \citet{lee96}. The number densities of H, H$_2$ and H$_3^+$ are 
shown on top of the pressure-temperature profile in Figure \ref{h2-abundance}.

The molecular hydrogen in the upper atmosphere of a brown dwarf in the presence of a strong UV field produced by a nearby white dwarf 
may become excited by absorbing photons in the lines of the $B ^1\Sigma_u-X ^1\Sigma_g^+$ and $C ^1\Pi_u-X ^1\Sigma_g^+$ bands. These molecules 
will rapidly transition into either the continuum ground-state, resulting in dissociation \citep{Field1966}, or will find themselves within the 
rovibrationally excited ground state. Molecular hydrogen in these excited states will be spontaneously de-excited 
into states progressively closer to the ground state. We determine the line intensity for IR fluorescence lines for rovibrationally excited H$_2$ within 
its electronic ground state, using \citet{Sternberg1989}:

\begin{equation}
F_{\nu}(v'j' \rightarrow vj) = \frac{h\nu}{4\pi} N(v'j') A(v'j'\rightarrow vj) g(\nu-\nu_0)
\end{equation}

where $\nu_0$ [Hz] is the transition frequency, $N(v'j')$ [cm$^{-2}$] is the column density of the excited species, $A$ [s$^{-1}$] is the 
rate of the transition from $v'j'\rightarrow vj$ and $g(\nu-\nu_0)$ is the line-shape, which we assume to be susceptible to Doppler broadening
only. The value of $N(v'j')$ can be determined 


 as in Eq. 2 from \citet{Sternberg1989}, using H$_2$ formation via three-body reactions \citep{Baulch1994}. A unitless scaling parameter in
this set of equations, $\chi$, represents the strength of the UV field at $\sim 100$ nm, and is given in Draine 
units \citep{Draine1978}. The value of $\chi$ is parameterised by \citet{Sternberg1989} (their Eq. A7):

\begin{equation}
 \chi = 3600 \Bigg( \frac{R}{R_{\odot}} \Bigg)^{\!\!2} \Bigg(\frac{r}{1 \; {\rm pc}} \Bigg)^{\!\!-2} \frac{1}{e^{14.4/T_4} - 1},
\end{equation}

where $T_4$ is the temperature of the white dwarf in units of $10^4$ K, $R$ is the radius of the white dwarf in units of solar radii and 
$r$ is the distance between the white dwarf and brown dwarf in units of parsecs. For WD0137, $T_{\rm eff} = 16500$ K, 
$R/R_{\odot} \approx 0.019$ and $r \approx 1.8 \times 10^{-6}$ pc. This yields an enormous value of $\chi = 6.5 \times 10^7$. A strong
UV  flux means that the rovibrational states of the molecular hydrogen will be highly populated, however if the field is very
high, the dissociation rates will also be large, and most of the hydrogen will be atomic. Because of the self-shielding of
molecular hydrogen, the UV pumping will nevertheless become efficient. When $\chi$ is large, H$_2$ fluorescence will become significant
deeper in the brown dwarf atmosphere. For $\chi \sim 10^7$, self-shielding does not become effective at protecting the molecular hydrogen 
until $\sim1$ mbar for our model atmosphere. The number density of UV-pumped H$_2$, $n({\rm H_2^*})$ 
[cm$^{-3}$], is shown in Figure \ref{h2-abundance}.

\subsubsection{ H$_{3}^{+}$ emission}

The molecular hydrogen can also be ionised by energetic particles and UV photons, forming H$_2^+$, which then reacts with molecular hydrogen
to form H$_3^+$. In the presence of strong external radiation, the gas in the upper atmosphere of the brown dwarf will become weakly ionised,
and we assume that free electrons will dominate the destruction of H$_3^+$. The H$_3^+$ abundance is solved using the chemical kinetics network,
albeit not self-consistently. Most importantly, the atmospheric dynamics from the model brown dwarf atmosphere are not coupled to the chemistry, and 
the effect of H$_3^+$ radiative cooling is not coupled to the thermal balance of the model atmosphere. Within these limitations, the chemistry calculations
predict that the majority of H$_3^+$ is between $10^{-6}$ bar (the low pressure limit of the model 
atmosphere), and $10^{-2}$ bar, with a peak abundance of $10^8$ cm$^{-3}$ (about 100 times the peak
abundance observed on Jupiter; see \citealt{Melin2005}, their Fig. 1).
When we calculate the line profile, we assume all the emitting energy levels are in local 
thermodynamic equilibrium (LTE), for simplicity; we take for our H$_{3}^{+}$ 
cross-sections the Exomol database values \citep{Neale1996, Hill2013}. Eventually, non-LTE models 
should be used, but LTE models are commonly used to model 
H$_{3}^{+}$ in Jupiter \citep{miller00} and non-LTE models suggest that, at least for 
Jupiter, the assumption of LTE for emitting energy levels accurately reproduces the overall emission
rate of H$_3^+$ \citep{Melin2005}. This assumption is sufficient for our purposes here.

\begin{figure}

\begin{center}

\scalebox{0.55}{\includegraphics{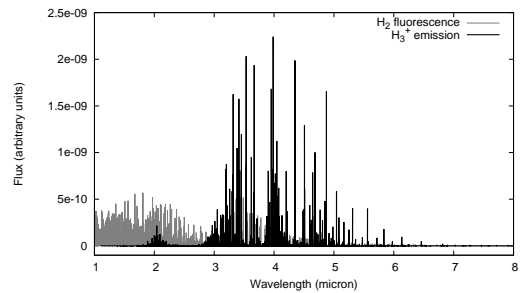}}

\caption{\label {h2} H$_{2}$ fluorescence (grey) and H$_{3}^{+}$ emission (black)
 spectra for WD0137-349B. The flux scale is in relative units, as we cannot calculate exactly how much 
emission would be expected without having models that fully incorporate photochemistry, the coupling to the thermal structure and the atmospheric dynamics. Additionally, the \citet{Sternberg1989}
approximation for the H$_{2}$ fluorescence has not been tested for high densities as yet.  We will address these issues in a future publication}.

\end{center}

\end{figure}

Both the modelled H$_{3}^{+}$ and H$_{2}$ emission spectra are shown in Figure \ref{h2}.

\section{Discussion}

WD0137-349B clearly has a large temperature difference between the irradiated and non-irradiated hemispheres. This difference is largest in the [4.5] micron band at 834 K, and is 710 K at [3.6] microns and 478 K in $K_{S}$, and the errors on these measurements are 89 K, 400 K, and 143 K, although as these errors are strongly correlated with the errors on the white dwarf T$_{\rm eff}$ and log g, they may be an overestimate. The error on the [3.6] micron band is large, due to the uncertainty associated with the minimum nightside flux.  In contrast, the exoplanet HD189733b has brightness temperatures of 1328 K in [3.6] and 1192 K in [4.5] microns, and a difference in 
day and night-side temperatures of  503 K at [3.6] microns, and 264 K at [4.5] microns  \citep{knutson12}. Wasp-12b is one of the most highly irradiated exoplanets  and is estimated to have a 3000 K  dayside and 1100 K nightside \citep{sing13}. While the temperature differences are not as extreme as for Wasp-12b, WD0137-349B is hotter than HD189773b on the dayside and colder on the nightside. 

Despite searches, there have been no detections of H$_{3}^{+}$ in either exoplanets or brown dwarfs to date (e.g. \citealt{goto}).  However, \citet{france13} have reported detections of H$_{2}$ fluorescence in planets around M dwarfs. This H$_{2}$ is detected in the Lyman $\alpha$ line in the UV. However, despite detecting the emission in four of the six M dwarf planetary systems studied, they are unable to confirm whether the H$_{2}$ has a chromospheric origin or is being emitted from the planet itself. It is evident that more detailed modelling is required, particularly in the UV wavelengths to predict whether H$_{2}$  is detectable within Lyman $\alpha$ in a white dwarf spectrum.

\citet{lammer03} argue that while $H_{3}^{+}$  is a common molecule that cools the thermosphere of  planets such as Jupiter \citep{miller00}, it may not have the same effect in exoplanets as those orbiting within 0.1 AU of their host star will reach temperatures high enough to dissociate H$_{2}$ ($\sim$ 10 000 K), preventing  its formation. This dissociation is also caused by the high UV irradiation at these close separations. \citet{koskinen07} claim that this is not entirely correct, as \citet{lammer03} neglected the cooling effect that H$_{3}^{+}$ produces, arguing instead that the presence of  $H_{3}^{+}$ can cool the atmospheres enough to generate more H$_{3}^{+}$.  \citet{yelle04} produced similar calculations and using 1D models determined that in some hot Jupiters the thermosphere can reach 10 000-15 000 K and still have cooling provided by H$_{3}^{+}$ at lower altitudes in the thermosphere, even though the H$_{2}$ is rapidly lost via thermal dissociation. Indeed, these models show that H$_{3}^{+}$ cooling has a significant impact on the temperature profiles of close-in exoplanets in spite of the low density of H$_{3}^{+}$ in the thermospheres of these exoplanets, compared to the H$_{3}^{+}$ densities in Jupiter's ionosphere.

\citet{koskinen07} get a slightly different result and also include radiative cooling and atmospheric circulation in their models. They note that dissociation occurs at a much lower 3000 K on a planet around a solar type star at $\sim$0.1 AU.  While WD0137-349B has a separation of 0.003 AU, the brightness and equilibrium temperatures are  much lower than 3000 K, because the white dwarf luminosity is lower than that of a solar-type star. 

There are a handful of brown dwarfs known to be irradiated by stellar companions (e.g. Wasp-30B \citealt{anderson}, Kelt-1b \citealt{beatty14}), however their equilibrium temperatures are in general too high to harbour  H$_{3}^{+}$.  For example, Kelt-1b \citep{beatty14}. It is estimated to have an brightness  temperature of $\sim$3000 K at [3.6] and [4.5] microns on the dayside, with a tentative detection of a temperature inversion suggesting a nightside cold trap indicating temperatures of $\sim$2000K. It orbits within 0.1 AU making  H$_{3}^{+}$  formation unlikely. This is supported by the [3.6]-[4.5] colour, which they note is identical to that of an isolated brown dwarf of similar temperature. 

While \citet{burleigh06} compared their nightside spectrum of WD0137-349B with a combined model of a white dwarf and an L8 dwarf in the near-IR, the $H$ band flux is much lower than this. This inconstancy  seems to indicate that in the mid-IR, a significant departure from a canonical brown dwarf spectrum is seen. This change is possibly due to energy transport around the brown dwarf from the heated to the unheated hemisphere. The measurements of the heated side however, do not seem to indicate a temperature inversion due to TiO as is seen in some exoplanets, but instead suggests a possible brightening in the $K$ and [4.5] micron bands, possibly due to H$_{3}^{+}$ and H$_{2}$ fluorescence. 

Phase resolved spectra of WD0137-349B in the $K$ band will allow some determination of how the spectrum alters with phase and whether species such as H$_{3}^{+}$ and effects such as H$_{2}$ fluorescence are occurring. L and M band spectroscopy from METIS on E-ELT or NIRSPEC from JWST could also be used confirm this effect. It is also clear that models of irradiated brown dwarfs and exoplanets need to be extended to include photochemistry and fluorescent effects.

\section{Summary}
We have observed the detached post-common envelope white dwarf+brown dwarf binary WD0137-349 in all four $Spitzer$ IRAC bands, obtaining an orbit at each wavelength as well as in the optical $R$, $I$ and $Z$ optical wavebands. Variability on the 116 minute orbit is seen at all wavelengths. The trend presented in \citet{casewell13} in which the aptitude of the peak-to-peak variability increases with wavelength is continued until 4.5 microns, after which the variability drops and remains constant. We calculate brightness temperatures and magnitudes for the brown dwarf alone and suggest that the tentative detection of a temperature inversion (based on brightness temperatures)  is possibly not real and may be due to some other effect such as the UV  irradiation of WD0137-349B creating photochemical effects in the atmosphere. These species are most luminous in the $K_{s}$ and [4.5] micron bands, thus potentially giving the appearance of a temperature inversion. 

\section{Acknowledgements}
We thank Derek Homier for his useful comments at the refereeing stage. SLC acknowledges support from the College of Science and Engineering at the University of Leicester. PBR and ChH highlight financial
    support of the European Community under the FP7 by an ERC starting grant. SPL is supported by STFC grant ST/J001589/1. This work is  based [in part] on observations made with the Spitzer Space Telescope, which is operated by the Jet Propulsion Laboratory, California Institute of Technology under a contract with NASA. Also based on observations made with ESO Telescopes at the La Silla Paranal Observatory under programme IDs 080.C-0587(A), 276.D-5014(A), 079.C-0683(B) and 080.C-0587(B).
This research has made use of NASA's Astrophysics Data System Bibliographic Services and the models hosted by Pierre Bergeron here: http://www.astro.umontreal.ca/$\sim$bergeron/CoolingModels.

\bibliographystyle{mn2e}

\bibliography{wd0137_bib}
\label{lastpage}

\end{document}